\documentclass[]{revtex4-1}
\usepackage[utf8]{inputenc}
\usepackage[margin=1in]{geometry}
\usepackage{charter}
\usepackage{fullpage}
\usepackage[lastpage,user]{zref}
\usepackage[colorlinks=false]{hyperref}
\usepackage{hyperref,amsmath,amsfonts}
\usepackage{braket}
\usepackage{graphicx}
\usepackage{multirow}
\usepackage[usenames]{color}
\usepackage{graphicx}
\usepackage{slashed}
\usepackage{mathptmx}

\begin{document}

\title{{Latin American network on electromagnetic effects in strongly interacting matter:} \\
Contribution to the update of the Latin American Strategy for High Energy, Cosmology and Astroparticle Physics}

\author{Alejandro Ayala$^{1}$, Ana Julia Mizher$^{2,3,4}$}
\affiliation{%
$^1$Instituto de Ciencias Nucleares, Universidad Nacional Aut\'onoma de M\'exico, Apartado Postal 70-543, CdMx 04510, Mexico.\\
$^2$ Instituto de F\' isica Te\'orica, Universidade Estadual Paulista, Rua Dr. Bento Teobaldo Ferraz, 271 - Bloco II, 01140-070 S\~ao Paulo, SP, Brazil.\\
$^3$Laboratório de Física Teórica e Computacional, Universidade Cidade de São Paulo, 01506-000, São Paulo, Brazil.\\
$^4$Centro de Ciencias Exactas and Departamento de Ciencias B\'asicas, Facultad de Ciencias, Universidad del B\'io-B\'io, Casilla 447, Chill\'an, Chile.}

\begin{abstract}

An accurate characterization of the quark-gluon plasma requires understanding of how electromagnetic effects affect the processes mediated by the strong force. All the scenarios in which the plasma emerges, either in nature or in the laboratory, involve strong electromagnetic fields. The early universe, compact astrophysical objects, or ultra-relativistic heavy-ion collisions harbor the most intense fields we know. Researches from the Latin America region have made a substantial contribution on this subject and the \lq\lq Latin American Network on Electromagnetic Effects in Strongly Interacting Matter" aims to cluster efforts to address open questions related to these systems, boosting collaborations and interaction among its members and connecting Latin American institutions with institutions from the rest of the world. In face of the upcoming experimental programs and new facilities, our mission is to bring together experimentalists, phenomenologists and theorists to better explore the properties of strongly interacting matter in the presence of intense electromagnetic fields. This document describes succinctly the recent contributions from researchers of the Latin American region to the subject, as well as our activities and perspectives for the future. \\

\noindent {\bf Thematic Areas: Quark gluon plasma, heavy-ion collisions, nuclear astrophysics, quantum field theory, condensed matter}

\end{abstract}

\maketitle

\vspace{0.4cm}

\section{Introduction}

The quark-gluon plasma (QGP) is a state of matter where quarks and gluons are deconfined, that is to say, not bound within hadrons, and resembles a strongly interacting liquid. In nature, it is believed to have been present in the early universe and currently to exist in the core of compact astrophysical objects, such as neutron stars. The way to produce this state of matter in laboratory experiments is by colliding heavy ions (HIC) at ultra-relativistic energies, thus inducing the necessary extreme conditions of temperature and density for a transition from a confined to a deconfined phase to occur. Understanding the formation and evolution of the plasma in the lab is therefore a way to obtain information about the very early stages of the universe. 

In addition to the large temperatures and densities achieved in heavy-ion reactions, non-central collisions produce the largest electromagnetic field strengths know in nature. Although the intensity drops down fast and, moreover, it is not clear whether the fields last long enough to induce a magnetization during the quark-gluon plasma phase, the models and simulations predict a significant intensity a few fermi after the beginning of the reaction.

Other systems in nature where the QGP can also be present are in turn influenced as well by strong electromagnetic fields. Magnetic fields have been observed from galaxy clusters to superclusters, and there is indirect evidence of a pervasive intergalactic magnetic field. Although their origin is unknown, magnetic fields have been observed in all astrophysical scales, which suggest their primordial origin, implying that instants after the Big-Bang, when the high energy and absence of thermal equilibrium kept matter in a deconfined state, a magnetic background was also present~\cite{Kronberg_1994}. On a less speculative arena, it is knwon that strong magnetic fields are present in the interior of neutron stars~\cite{magstars_review, pulsars,Harding_2006}. These are super-dense objects formed when a massive star runs out of fuel and collapses. Their core may be composed by hadrons, or, in the case the density overcomes a critical value, it may be composed by deconfined quarks, or a combination of both. Their typical mass is of the order of $1-2$ solar masses and have radii of the order of 10 km. Therefore, they are extremely dense objects. The high densities do not favor the formation of nuclei in their core and lighter particles are more likely to appear. Protons and neutrons are expected to be found in the interior of neutron stars and, for densities larger than a critical value, these particles may undergo a phase transition to a deconfined state. In this case the quark-gluon plasma is formed and the object is called a quark star. 

Neutron stars (regardless of whether their interior is made either of quarks or hadrons) possess large magnetic field strengths on their surface which is believed to originate from the core. Observations indicate a field strength in the range $10^{11} - 10^{15}$G on the surface~\cite{magstars_review, pulsars,Harding_2006,NS} and it is estimated that the core possibly harbors field strengths a few orders of magnitude larger~\cite{PhysRevLett.112.171102}. Such fields influence the matter equation of state and the general relativistic solutions. Therefore, these fields affect the mass-radius relation and must be taken into account to determine if the currently developed models must be kept or ruled out when compared to observations. 

The study of electromagnetic effects on strongly interacting matter has been a subject of intense activity during the last decade. 
Besides becoming a subject in traditional events like {\it Quark Matter} and {\it Extreme QCD} - that cover heavy-ion physics and QCD in extreme conditions - the topic has motivated focused workshops, such as the \href{https://indico.ihep.ac.cn/event/16043/}{``International Conference on Chirality, Vorticity and Magnetic Field in Heavy Ion Collisions"}. that has been organized periodically since 2017; the \href{https://www.ictp-saifr.org/eesm2022/}{``Workshop on Electromagnetic Effects in Strongly Interacting Matter} in 2022; the \href{https://www.ectstar.eu/workshops/strongly-interacting-matter-in-extreme-magnetic-fields/}{``Strongly Interacting Matter in Extreme Magnetic Fields"} in September 2023; and more recently the \href{https://indico.nucleares.unam.mx/event/2216/}{``First Latin American Workshop on Electromagnetic Effects in QCD} in July 2024.

\section{Short description of recent contributions from Latin American researchers}

As an example of very recent contributions to the field made by researchers from the Latin American region, hereby we present a small summary of works published during the last year (2024) on the subject of the influence of magnetic fields on the properties of strongly interacting matter.

\subsection{Anisotropic gluon pressure influenced by magnetic fields during pre-equilibrium in relativistic heavy-ion collisions}

In relativistic heavy-ion collisions, pre-equilibrium is characterized by the existence of strong color fields which are liberated after the Glasma is shattered at the beginning of the reaction. In recent times, it has also been realized that other relevant players during pre-equilibrium are the strong electromagnetic fields produced in peripheral collisions~\cite{STAR:2023jdd}. A magnetic field also produces an anisotropic pressure between the parallel and perpendicular directions, with respect to the magnetic field, and thus its contribution to the overall anisotropy during pre-equilibrium needs to be accounted for. In a recent work~\cite{Ayala:2024jvc}, it has been shown that the magnetic field-induced pressure contributes to tame the pre-equilibrium anisotropy and speeds up the road towards thermalization.

\subsection{Excess of photon and $v_2$ from gluon fusion and splitting in relativistic heavy-ion collisions}

In semicentral relativistic heavy-ion collisions, magnetic fields provide a source of electromagnetic radiation that at the same time induce a natural anisotropic emission, and thus contribute to $v_2$, without the need to link its strength to the flow properties of the system. Since the field intensity peaks at the early stages of the collision, where gluons dominate over quarks, it is important to study the emission of electromagnetic radiation induced by gluon-driven processes. Two main channels are the photon production from gluon fusion and splitting. An accurate estimate of the contribution from these processes requires knowledge of the general structure of the two-gluon one-photon vertex in the presence of a magnetic field. In a recent work~\cite{Ayala:2024ucr} this structure was found, and the result used to compute the explicit one-loop two-gluon one-photon vertex expression for the intermediate field strength, with respect to the photon transverse momentum.

\subsection{Meson masses in a magnetic field}

Recall that for a Lorentz-invariant system, the mass corresponds to the rest energy of a given particle, which can then be obtained from the pole of the propagator when the particle three-momentum $\vec{q}$ is taken to zero. This is dubbed the ``{\it pole mass}''.  Notice that if, instead, the zeroth component of the particle four-momentum $q_0$ is taken first to zero, we obtain the ``{\it screening mass}''. The screening mass squared can be identified as the negative of the particle magnitude of its three-momentum squared. In a system where Lorentz symmetry is unbroken, the pole and screening masses coincide. However, when Lorentz symmetry is broken, the above described limiting procedures do not yield the same values. The name screening mass stems from the analysis in linear response theory when studying the influence of static external fields on a thermal medium. Because of the static nature of the external field, its screening within the medium is controlled by the system's response function in the limit $q_0=0$, $\vec{q}\to 0$. The inverse of the screening mass corresponds to the screening or Debye length. When the system is immersed in a magnetic field, the breaking of Lorentz symmetry happens in the spatial directions, giving rise to distinct dispersion properties for particles moving in the transverse or the longitudinal directions with respect to the field orientation. Thus, in addition to studying the magnetic-field-induced modifications of the pole mass, one can also study the corresponding longitudinal and transverse screening masses. In Ref.~\cite{Coppola:2023mmq}, the authors study the pole mass spectrum of the light pseudoscalar and vector mesons in the presence of an
external uniform magnetic field $B$, considering the effects of the mixing with the axial vector meson sector. The analysis is performed within a two-flavor NJL-like model which includes isoscalar and
isovector couplings together with a flavor mixing ’t Hooft-like term. For neutral pion
masses it is shown that the mixing with axial vector mesons improves previous theoretical results,
leading to a monotonic decreasing behavior with B that is in good qualitative agreement with LQCD calculations. For charged pions, it is seen that the mixing softens the enhancement of their mass with $B$. As a consequence,
the energy becomes lower than the one corresponding to a pointlike pion, improving the agreement
with LQCD results. In Ref.~\cite{Ayala:2023llp}, the authors use the linear sigma model with quarks to study the magnetic field-induced modifications on the longitudinal screening mass for the neutral pion at one-loop level. They find that to obtain a reasonable description for the behavior with the field strength, one needs to
account for the magnetic field dependence of the particle pole masses and that the couplings
need to decrease fast enough with the field strength to then reach constant and smaller values as
compared to their vacuum ones. The results illustrate the need to treat the magnetic corrections
to the particle masses and couplings in a self-consistent manner, accounting for the back reaction of
the field effects for the magnetic field dependence of the rest of the particle species and couplings
in the model. In Ref.~\cite{Hernandez:2025inu}, the authors screening mass of the neutral rho-meson in the presence of strong magnetic fields
using the Kroll-Lee-Zumino model. They show that the zero and perpendicular modes
exhibit a monotonically increasing behavior with the magnetic field strength, whereas the parallel
mode remains essentially constant. Their findings provide new insights into the behavior of vector
mesons in strongly magnetized media.

\subsection{Fluctuating magnetic background}

Almost all calculations of magnetic field effects on particle properties are carried out assuming a constant magnetic background or at most an adiabatic change with time. Nevertheless, a more realistic situation may be described if the magnetic field is considered to possess
stochastic fluctuations, described as white noise. Implications of this formulations are studied in Ref.~\cite{Castano-Yepes:2024ctr} for the case of photons and in Ref.~\cite{Castano-Yepes:2024ltr} for the case of fermions. In the former, it is found that the fluctuations of the background give rise to a photon magnetic mass (or, in an extension of the ideas to the case of QCD, also to a gluon magnetic mass). In the latter it is found that, while a uniform magnetic field already breaks Lorentz
invariance, inducing the usual separation into two orthogonal subspaces (perpendicular and parallel with respect
to the field direction), the presence of magnetic noise further breaks the remaining symmetry, thus leading to distinct
spectral widths associated with fermion and anti-fermion, and their spin projection in the quasi-particle picture.

\subsection{The quark anomalous magnetic moment }

The ordinary anomalous magnetic moment (AMM) of quarks may be calculated perturbatively, via loops in Feynman diagrams. In addition, nonperturbative calculations that account for the dynamical chiral symmetry breaking, provide an important contribution to the AMM, specially for massless fermions, for which perturbative calculations predict a vanishing AMM. In \cite{Fraga:2024klm}, the deviation of the anomalous magnetic moment of quarks when those are in a strong magnetic background is obtained, considering perturbative 1-loop corrections for the photon-quark-antiquark vertex, in the case where the magnetic field is the largest scale. It is shown that an asymmetry between spin up and spin down components of the anomalous correction emerges, suggesting possible experimental observation. The approach has in common with nonperturbative calculations the fact that the deviation depends on chiral symmetry breaking, vanishing in the chiral limit. 

On the other hand, in \cite{Tavares:2023oln,Tavares:2024myk} the influence of the anomalous magnetic moment on the nature of the QCD phase transition is calculated. Nonperturbative features are mimicked by the use of an effective model, a thermo-magnetic Nambu-Jona-Lasinio with a Pauli term, responsible for the coupling to the AMM of the quarks. It is found that the order of the phase transition depends on the regularization scheme and that in the presence of a magnetic background, the magnetic field independent regularization (MFIR) washes out the first order phase transition usually found when applying other methods, yielding to a crossover consistent with lattice calculations. 

\subsection{Magnetic corrections to the strong coupling and form factors}

Lattice QCD have been for over a decade obtaining quantitative information about strongly interacting systems in a magnetic background. Characterization of the phase structure of QCD and modified hadron masses, when a magnetic background is present, are some of the main results obtained recently by lattice simulations in this topic. Regarding the phase structure, when the system is exposed to a high temperature simultaneously with a magnetic background, simulations predict a decrease of the critical temperature of the chiral and deconfinement transition, a phenomenon known as inverse magnetic catalysis (in opposition to the magnetic catalysis that takes place in the vacuum where the magnetic field acts to enhance the chiral symmetry breaking). This behavior is interpreted as coming from a back reaction of the sea quarks to the field. An effective way to encode this mechanism in theories and effective models describing the strong interaction is to consider that the intensity of the couplings depends on the magnitude of the magnetic field. This procedure has been successfully explored by several groups in a way that currently effective approaches manage to reproduce lattice results. 


Following this reasoning, it is very important in all the approaches to understand how a magnetic background affects the coupling between particles, being them fundamental particles or bound states. In \cite{Fernandez:2024tuk}, the gluon-quark and the three-gluons vertices are evaluated considering 1-loop magnetic corrections within the lowest Landau level approach, valid for extremely large magnetic fields. It is shown that the effective coupling grows when the field increases. Besides loop calculations in QCD, effective models may also be explored to obtain non-perturbative insight on aspects like the thermodynamical properties of deconfined quarks and gluons. In \cite{Valenzuela-Coronado:2024ewi}, the authors propose a thermo-magnetic bag model in order to estimate the pressure of a magnetized quark-gluon plasma. To achieve this goal, they introduce the information about the magnetic field through the masses, applying a coupling that depends on the intensity of the magnetic field.


In addition to the coupling, form factors will likely to be affected by the presence of a magnetic field. In \cite{Braghin:2023ykz}, the authors estimate magnetic corrections to a Yukawa potential, modeling the quark-quark interaction via a pion exchange. Magnetic corrections to the coupling, propagators and form factors are taken into account to estimate the potential. It is found that the magnetic contributions may be isotropic or anisotropic, depending on the masses considered for the pion and the quarks.

Magnetic corrections to the coupling of strongly interacting matter also must be considered in the context of nuclear astrophysics. In Ref.~\cite{Dominguez:2023eyg}, QCD finite energy sum rules are applied to calculate the axial-vector coupling constant in a magnetic background and in a dense medium. This configuration is interesting to explore features of magnetars, since it is estimated that these objects present a magnetic field of the order of $10^{15}$G  on their surface and that the intensity of this field can reach a few orders of magnitude more in their core. It was found that the coupling decreases with baryonic density, while it is insensitive to the magnetic field, for the values of intensity considered in the paper.

\subsection{Perturbative QCD}

Due to asymptotic freedom, at low energies the coupling of QCD is large and perturbative calculations do not apply. However the regime of very high energies can be explored as an interesting limit, that must be reproduced in non-perturbative approaches. In \cite{Fraga:2023cef}, the authors explore perturbative QCD in a magnetic background, calculating relevant quantities, such as the chiral condensate and the susceptibility up to 2-loops, in a regime where the magnetic field is the largest scale. They also study the convergence of the perturbative series for the pressure for different choices of renormalization scale in the running coupling. In a complementary manner, in \cite{Fraga:2023lzn} the same authors calculate the pressure within perturbative QCD in the regime of cold and dense matter, relevant for astrophysical objects such as magnetars. The calculation is again performed up to 2-loops and the regime is such that the magnetic field is again the larger scale. The equation of state for pure quark magnetars is obtained and the results provide constraints on the behavior of the maximum
mass and associated radius from perturbative QCD.

\subsection{Electric field}

In addition to the strong magnetic field generated in non-central heavy-ion collisions, an electric field is generated in the case where the collision involves different species of ions, such as Au-Cu collisions. This is due to the imbalance in the number of protons associated to each nucleus. Electric and magnetic fields are generated perpendicular to each other: while the magnetic field appears perpendicular to the direction of the reaction plane, the electric field is longitudinal to the direction of the displacement of the beam. It is therefore interesting to study the effects of an electric field on the properties of hadrons. In \cite{Cadiz:2024rzs} the \ensuremath{\pi}-\ensuremath{\pi} scattering lengths are calculated in the case an electric field is present. The limit of strong field must be taken carefully, since  Schwinger instabilities associated to pair productions of pions could appear. Therefore, in this work the authors restrict themselves to the weak field limit. Also exploring effects of electric fields on charged particles, in \cite{Tavares:2024edx} the authors study the symmetry breaking and restoration behavior of a self-interacting charged scalar field theory under the influence of a constant electric field and finite temperature. The phase diagram is explored, determining critical temperatures and the nature of the phase transitions.

\subsection{Condensed matter}

The emergence of Weyl and Dirac materials have opened a window to explore in condensed matter phenomena that in principle belong to the realm of high energy systems. This is possible since the charge carriers in this kind of material exhibit a relativistic-like behavior, obeying Dirac equation, as an effective result of their interaction with the underlying lattice. In particular, transport phenomena predicted to take place in the quark-gluon plasma have been explored in a few configurations of Weyl materials. Due to this relativistic-like behavior, QED in $(3+1)$ or $(2+1)$ dimensions is used as a continuum limit of tight-binding approaches, depending on the dimensionality of the material. For planar material such as graphene, an issue occurs when the electrons in the material interact with external fields. A complementary approach is needed, since external fields do not live in the same dimensionality as the fermions. Reduced QED (RQED), also known as Pseudo-QED (PQED), is a mixed dimension theory developed to treat interactions where different particles are constrained to different dimensions. In \cite{Mizher:2024zag}, the authors explore the Fried-Yennie gauge applying it to RQED. This is a covariant gauge, specially convenient in this case since it induces a cancellation involving the gauge parameter that simplifies the infrared sector of the theory, without the need of inclusion of a photon mass, usually required when working in other gauges.

\section{Perspectives}

We emphasize that the topics briefly discussed in this contribution are a non-exhaustive collection of the work done by members of our network published in 2024 and submitted in 2025. As discussed in the workshop~\href{https://www.ectstar.eu/workshops/strongly-interacting-matter-in-extreme-magnetic-fields/}{``Strongly Interacting Matter in Extreme Magnetic Fields"}, it is necessary to apply the important theoretical progress acquired in the last decade to study observables in strongly interacting systems. The network will turn its efforts to this direction in 2025. As a first initiative taken in 2024, we organized the mini-course "Hydrodynamics models of Heavy-Ion Collisions" given by Prof. Gabriel Denicol from the Universidade Federal Fluminense, available at https://www.youtube.com/watch?v=vXxSixbbJvs.

The members of the network have been meeting regularly, being it during the monthly virtual seminar or during the in-person workshops organized yearly. We plan to keep organizing the seminars, inviting prominent researchers working on the topic. We aim to cover specially the techniques not covered by our members, important recent papers and experimental results. 

In 2025, the {\it IX International Conference on Chirality, Vorticity and Magnetic Fields in Quantum Matter} will take place in São Paulo organized by 2 of our members https://www.ictp-saifr.org/ccvmfqm2025/, joining researchers from Latin America and abroad, from theory and experiment, to discuss related topics. This is the most important meeting in the subject and Brazil was chosen as the 2025 host due to the live activity of Latin America in the field.

\section{Network description}

\vspace{0.5cm}

\noindent {\bf Network webpage:} \href{https://www.ictp-saifr.org/laqcd/}{https://www.ictp-saifr.org/laqcd/}

\vspace{0.5cm}

\noindent {\bf Network youtube channel:} \href{https://www.youtube.com/@LatinAmericanEM-QCD-og7gf}{https://www.youtube.com/@LatinAmericanEM-QCD-og7gf}

\vspace{0.5cm}

\noindent {\bf Participants} 

\noindent Currently, we have 27 members from 20 institutions, belonging to 6 Latin American countries. In addition we have 4 external members. Our members are distributed in the following way: 

\vspace{0.3cm}

{\bf Brazil} (15 participants)

\begin{itemize}

\item   Universidade Cidade de São Paulo (Unicid)
\item	Instituto de Física Teórica - (IFT-UNESP)
\item	Universidade Estadual de Campinas (UNICAMP)
\item	Universidade Federal do Rio de Janeiro (UFRJ)
\item   Universidade Estadual do Rio de Janeiro (UERJ)
\item	Universidade Federal de Santa Catarina (UFSC)
\item	Universidade Federal de Santa Maria (UFSM)
\item	Universidade Federal do Rio Grande do Sul (UFRGS)
\item   Universidade de São Paulo (USP)
\item   Universidade Federal de Goiás (UFG)
\item   Instituto Tecnológico da Aeronáutica (ITA)

\end{itemize}

\vspace{0.3cm}

{\bf Argentina} (2 participants)

\begin{itemize}
    \item Department of Theoretical Physics, Comisión Nacional de Energía Atómica (CNEA)

\end{itemize}

\vspace{0.3cm}

{\bf Chile} (3 participants)

\begin{itemize}
    \item Pontificia Universidad Católica de Chile (PUC-Chile)
 \item Universidad del Bio-Bio (UBB)

\end{itemize}

\vspace{0.3cm}

{\bf Cuba} (1 participants)

\begin{itemize}
    \item Universidad de La Habana

\end{itemize}

\vspace{0.3cm}

{\bf Colombia} (1 participants)

\begin{itemize}
    \item Universidad del Valle
    
\end{itemize}

\vspace{0.3cm}

{\bf Mexico} (5 participants)

\begin{itemize}
    \item Universidad Nacional Autónoma de Mexico (UNAM)
\item	Universidad Autónoma Metropolitana (UAM)
\item	Universidad de Colima (UCol)
\item	Universidad Michoacana de San Nicolás Hidalgo (UMSNH)

\end{itemize}

{\bf External} (4 participants)

\begin{itemize}
    \item Kent State University
    \item Universitat de Barcelona
    \item University of York
    \item University of Illinois

\end{itemize}

\section{Funding}

Several members of our network receive individual funding from their respective countries, by agencies CNPq, FAPESP, FAPERJ, Serrapilheira, Conicyt-Chile, Conicet-Argentina and CONAHCyT-Mexico. All the students count on fellowships from the same agencies.  The events have been supported by ICTP-SAIFR, CLAF and IANN-QCD. The initiative is part of the Jovem Pesquisador FAPESP grant number 2023/08826-7, received by the coordinator Ana Mizher. 

\pagebreak

\bibliographystyle{unsrt}
\bibliography{bibliography}

\begin{thebibliography}{10}

\bibitem{Kronberg_1994}
P~P Kronberg.
\newblock Extragalactic magnetic fields.
\newblock {\em Reports on Progress in Physics}, 57(4):325, apr 1994.

\bibitem{magstars_review}
G.~Chanmugam.
\newblock Magnetic fields of degenerate stars.
\newblock {\em Annual Review of Astronomy and Astrophysics}, 30(1):143--184, 1992.

\bibitem{pulsars}
R.~N. Manchester and J.~H. Taylor.
\newblock {\em {Pulsars}}.
\newblock W. H. Freeman, San Francisco, 1977.

\bibitem{Harding_2006}
Alice~K Harding and Dong Lai.
\newblock Physics of strongly magnetized neutron stars.
\newblock {\em Reports on Progress in Physics}, 69(9):2631, aug 2006.

\bibitem{NS}
P.~Haensel, A.~Potekhin, , and D.~Yakovlev.
\newblock {\em Neutron Stars 1: Equation of State and Structure}.
\newblock Springer, New York, 2006.

\bibitem{PhysRevLett.112.171102}
K.~Makishima, T.~Enoto, J.~S. Hiraga, T.~Nakano, K.~Nakazawa, S.~Sakurai, M.~Sasano, and H.~Murakami.
\newblock Possible evidence for free precession of a strongly magnetized neutron star in the magnetar 4u $0142+61$.
\newblock {\em Phys. Rev. Lett.}, 112:171102, Apr 2014.

\bibitem{STAR:2023jdd}
M.~I. Abdulhamid et~al.
\newblock {Observation of the electromagnetic field effect via charge-dependent directed flow in heavy-ion collisions at the Relativistic Heavy Ion Collider}.
\newblock {\em Phys. Rev. X}, 14(1):011028, 2024.

\bibitem{Ayala:2024jvc}
Alejandro Ayala and Ana~Julia Mizher.
\newblock {Influence of magnetic field-induced anisotropic gluon pressure during pre-equilibrium in heavy-ion collisions: A faster road toward isotropization}.
\newblock {\em Phys. Rev. D}, 110(11):L111501, 2024.

\bibitem{Ayala:2024ucr}
Alejandro Ayala, Santiago Bernal-Langarica, Jorge Jaber-Urquiza, and Jos\'e~Jorge Medina-Serna.
\newblock {Two-gluon one-photon vertex in a magnetic field and its explicit one-loop approximation in the intermediate field strength regime}.
\newblock {\em Phys. Rev. D}, 110(7):076021, 2024.

\bibitem{Coppola:2023mmq}
M\'aximo Coppola, Daniel Gomez~Dumm, Santiago Noguera, and Norberto~N. Scoccola.
\newblock {Masses of magnetized pseudoscalar and vector mesons in an extended NJL model: The role of axial vector mesons}.
\newblock {\em Phys. Rev. D}, 109(5):054014, 2024.

\bibitem{Ayala:2023llp}
Alejandro Ayala, Ricardo L.~S. Farias, L.~A. Hern\'andez, Ana~Julia Mizher, Javier Rend\'on, Cristian Villavicencio, and R.~Zamora.
\newblock {Magnetic field dependence of the neutral pion longitudinal screening mass in the linear sigma model with quarks}.
\newblock {\em Phys. Rev. D}, 109(7):074019, 2024.

\bibitem{Hernandez:2025inu}
Luis~A. Hern\'andez, Juan~D. Mart\'\i{}nez-S\'anchez, and R.~Zamora.
\newblock {Screening rho-meson mass in the presence of strong magnetic fields}.
\newblock 2 2025.

\bibitem{Castano-Yepes:2024ctr}
Jorge~David Casta\~no Yepes and Enrique Mu\~noz.
\newblock {Exploring magnetic fluctuation effects in QED gauge fields: Implications for mass generation}.
\newblock {\em Phys. Rev. D}, 109(5):056007, 2024.
\newblock [Erratum: Phys.Rev.D 109, 119903 (2024)].

\bibitem{Castano-Yepes:2024ltr}
Jorge~David Casta\~no Yepes and Enrique Mu\~noz.
\newblock {Fermion self-energy and effective mass in a noisy magnetic background}.
\newblock {\em Phys. Rev. D}, 110(5):056003, 2024.

\bibitem{Fraga:2024klm}
Eduardo~S. Fraga, Leticia~F. Palhares, and Cristian Villavicencio.
\newblock {Quark anomalous magnetic moment in an extreme magnetic background from perturbative QCD}.
\newblock {\em Phys. Rev. D}, 109(11):116018, 2024.

\bibitem{Tavares:2023oln}
William~R. Tavares, Sidney~S. Avancini, Ricardo L.~S. Farias, and Rafael~P. Cardoso.
\newblock {Artificial first-order phase transition in a magnetized Nambu\textendash{}Jona-Lasinio model with a quark anomalous magnetic moment}.
\newblock {\em Phys. Rev. D}, 109(1):016011, 2024.

\bibitem{Tavares:2024myk}
William~R. Tavares, Rodrigo~M. Nunes, Sidney~S. Avancini, and Ricardo L.~S. Farias.
\newblock {The influence of quark anomalous magnetic moment in the Nambu\textendash{}Jona-Lasinio model with different regularizations}.
\newblock {\em Astron. Nachr.}, 345(2-3):e230168, 2024.

\bibitem{Fernandez:2024tuk}
Gabriela Fern\'andez, Luis~A. Hern\'andez, and R.~Zamora.
\newblock {Magnetic corrections to the QCD coupling: Strong field approximation}.
\newblock {\em Phys. Rev. D}, 110(1):014016, 2024.

\bibitem{Valenzuela-Coronado:2024ewi}
Paulina~Fernanda Valenzuela-Coronado, Maria~Elena Tejeda~Yeomans, and Jose Torres-Arenas.
\newblock {A thermo-magnetic bag model for the quark-gluon plasma}.
\newblock {\em Rev. Mex. Fis.}, 70(2):021201, 2024.

\bibitem{Braghin:2023ykz}
Fabio~L. Braghin, Marcelo Loewe, and Cristian Villavicencio.
\newblock {Yukawa potential under weak magnetic field}.
\newblock {\em Phys. Rev. D}, 109(3):034014, 2024.

\bibitem{Dominguez:2023eyg}
C.~A. Dominguez, M.~Loewe, C.~Villavicencio, and R.~Zamora.
\newblock {Nucleon axial coupling constant in a magnetar environment}.
\newblock {\em Nucl. Part. Phys. Proc.}, 343:43--47, 2024.

\bibitem{Fraga:2023cef}
Eduardo~S. Fraga, Let\'\i{}cia~F. Palhares, and Tulio~E. Restrepo.
\newblock {Hot perturbative QCD in a very strong magnetic background}.
\newblock {\em Phys. Rev. D}, 108(3):034026, 2023.

\bibitem{Fraga:2023lzn}
Eduardo~S. Fraga, Let\'\i{}cia~F. Palhares, and Tulio~E. Restrepo.
\newblock {Cold and dense perturbative QCD in a very strong magnetic background}.
\newblock {\em Phys. Rev. D}, 109(5):054033, 2024.

\bibitem{Cadiz:2024rzs}
R.~Cadiz, M.~Loewe, and R.~Zamora.
\newblock {Electric corrections to \ensuremath{\pi}-\ensuremath{\pi} scattering lengths in the linear sigma model}.
\newblock {\em Phys. Rev. D}, 109(11):116004, 2024.

\bibitem{Tavares:2024edx}
William~R. Tavares, Rudnei O.~Ramos, Ricardo L.~S. Farias, and Sidney~S. Avancini.
\newblock {Charged scalars at finite electric field and temperature in the optimized perturbation theory}.
\newblock {\em Phys. Rev. D}, 110(11):116024, 2024.

\bibitem{Mizher:2024zag}
Ana Mizher, Alfredo Raya, and Kh\'epani Raya.
\newblock {Fried-Yennie Gauge in Pseudo-QED}.
\newblock {\em Entropy}, 26(2):157, 2024.

\end{thebibliography}

\end{document}